\documentclass[preprint, trackchanges]{aastex631}

\begin{document}

\title{NuSTAR Analysis of the 2024 Periastron Passage of TeV Binary PSR B1259-63}

\correspondingauthor{Oliver Roberts}
\email{oroberts@usra.edu}

\author[0000-0002-7150-9061]{Oliver J. Roberts}
\affiliation{Science and Technology Institute, Universities Space and Research Association, Huntsville, AL 35805, USA.}

\author[0000-0002-3638-0637]{Philip Kaaret}
\affiliation{NASA Marshall Space Flight Center, Huntsville, AL 35812, USA}

\author[0000-0001-7163-7015]{M. Lynne Saade}
\affiliation{Science and Technology Institute, Universities Space and Research Association, Huntsville, AL 35805, USA.}

\author[0000-0002-4945-5079]{Chien-Ting Chen}
\affiliation{Science and Technology Institute, Universities Space and Research Association, Huntsville, AL 35805, USA.}

\author[0000-0003-4420-2838]{Steven R. Ehlert}
\affiliation{NASA Marshall Space Flight Center, Huntsville, AL 35812, USA}

\author[0000-0001-9200-4006]{Ioannis Liodakis}
\affiliation{NASA Marshall Space Flight Center, Huntsville, AL 35812, USA}
\affiliation{Institute of Astrophysics, Foundation for Research and Technology-Hellas, GR-70013 Heraklion, Greece}

\author[0000-0002-5270-4240]{Martin C. Weisskopf}
\affiliation{NASA Marshall Space Flight Center, Huntsville, AL 35812, USA}

\begin{abstract}

PSR B1259-63 is a well studied TeV binary, with an energetic pulsar in orbit around a Be star. Using NuSTAR observations during the 2024 passage of the pulsar through the circumstellar disk, we find the spectrum to be the most energetic ($\Gamma$ = 1.5) around 27 days after periastron, during the first of two variable, short-term emission episodes of a contemporaneous GeV flare. We discuss the variability in the X-ray flux and the hardening of the spectrum with time, and in the context of previous observations and what that means for the competing energy loss and acceleration timescales.    
\end{abstract}


\section{Introduction} \label{sec:intro}

PSR B1259-63 (hereafter B1259), is a radio pulsar with a spin period of P = 47.76~ms and spin-down power of approximately 8 $\times$ 10$^{35}$ erg/s~\citep{Johnston1994}. The pulsar is in a binary system with a massive, emission-line star with spectral type O9.5Ve (often referred to as a Be star), approximately 2.6~kpc away~\citep{MillerJones2018}. Optical spectra of the star, LS 2883, provide evidence for a dense equatorial outflow (isotropic wind), that forms a circumstellar disk, purportedly inclined with respect to the orbital plane~\citep{Dubus2013, Melatos1995}. The shortest distance between B1259 and LS 2883 is about 0.9 A.U.~\citep{Negueruela2011}, believed to be the size of the equatorial disk~\citep{Johnston1992}. The pulsar is in a highly eccentric orbit (e $\sim$ 0.87, P = 1236.724526 days or $\sim$ 3.38 years) around LS 2883, plunging through its disk twice, several weeks before and after periastron. The disappearance of radio pulsations around periastron accurately signals the pulsar's transit through the disk~\citep{Johnston1999}, accompanied by brightening in the X-ray band~\citep{Chernyakova2009}. This brightening in the X-ray band is widely accepted to be due to the interaction between the energetic pulsar wind and stellar outflow which results in highly energetic particles ($e^{-}e^{+}$ pairs) being efficiently accelerated in a time-varying shock~\citep{Tavani1997}. The resulting synchrotron radiation includes the radio and X-ray/soft gamma ray band, with the X-ray spectrum well-modeled by a Power Law (PL)~\citep{Grove1995,Chernyakova2009,Chernyakova2015}.

GeV flares were discovered during the 2011 periastron passage of B1259~\citep{Abdo2011,Tam2011} and have been observed in the periastron passages that followed~\citep{Tam2015,Tam2018,Chernyakova2020,Chernyakova2024,Chernyakova2025}. The 2024, 2014 and 2011 GeV flares occur at roughly the same orbital phase (starting and ending about 20 to 80 days post-periastron, respectively), with contemporaneous X-ray activity reported during the 2024 and 2014 GeV flares~\citep{Chernyakova2015,Tam2015,Chernyakova2025}. The onset of the 2017 and 2021 GeV flares were delayed (roughly 20 and 30 days relative to the 2011, 2014 and 2024 GeV flares, respectively~\citep{Chang2021}), with short-lived flaring occurring on timescales down to several hours during the 2017 flare~\citep{Tam2018}. It has been postulated that the delay of these GeV flares in 2021 and 2017 could be due to a slow precessing warped Be equatorial disk~\citep{Chang2021}. However, this scenario is not fully consistent with multi-wavelength observations from the 2024 periastron passage~\citep{Chernyakova2025} and therefore, the reasoning for the delay of the GeV flares during the 2017 and 2021 periastron passages remains ambiguous.

While the mechanism of the GeV flare is unclear~\citep{Dubus2013}, several models have been put forward which mostly involve the acceleration of electrons in a shock region between the pulsar wind and the stellar wind, from which synchrotron radiation is produced~\citep{Tavani1997}. Up-scattered photons from the circumstellar disk may also produce inverse-Compton radiation~\citep{Kirk1999,Kong2011,Mochol2013}. The interaction between the stellar disk and the pulsar has also been suggested to play a role in the production of GeV flares, with the ``clumpiness" or density, and size of the stellar disk purported to account for some of the characteristics observed for some of the more recent GeV flaring behavior~\cite{Chernyakova2024,Chernyakova2025}.

In this paper, we present the Nuclear Spectroscopic Telescope Array (hereafter NuSTAR;~\cite{Harrison2013}) results as part of a contemporaneous observing campaign with the Imaging X-Ray Polarimetry Explorer (IXPE;~\cite{Weisskopf2022}) about 15 days after periastron. We discuss the NuSTAR observations and analysis in Section~\ref{sec:analobs}, before presenting the results in Section~\ref{sec:res}. We compare and discuss these results in the context of previous observations of the source in Section~\ref{sec:disc}. 

\section{NuSTAR Observation and Analysis}\label{sec:analobs}

NuSTAR is a high-energy (3-79~keV) imaging telescope, launched on June 13, 2012~\citep{Harrison2013}. Photons from B1259 were focused using NuSTAR's two Wolter-I telescopes (each with a focal length of roughly 10~m), and subsequently collected in two Focal Plane Modules (FPMs A and B). The co-aligned X-ray telescopes provide imaging at resolutions of 58 arc-minutes (half power diameter) and 18 arc-minutes (FWHM). The Field of View (FoV) is 12 x 12 arc-minutes$^{2}$ at 10 keV. 

The live times of the four NuSTAR observations of B1259 are 24.0, 23.7, 25.2 and 28.2~ks (total exposure of 101.1~ks), with observation IDs; 3100200-3002, 3100200-3004, 3100200-3006 and 3100200-3008, respectively (hereafter, 3002, 3004, 3006 and 3008). The average distances between the J2000 coordinates of B1259 and the optical axis during observations 3002, 3004, 3006 and 3008 were 2.0, 1.7, 1.4, 1.6 arc-minutes for FPM A, and 2.6, 2.3, 2.0, 2.2 arc-minutes for FPM B, respectively. These NuSTAR observations are summarized in Table~\ref{tab:1}.

 \begin{table}[h]
    \centering
    \begin{tabular}{|c|c|c|c|c|c|c|}
    \hline
      ObsIDs & Start Time  & t-$\tau_{\rm p}$  &  Exposure  &  Flux (1-10 keV)  & $\Gamma$ & $\chi^{2}$/dof \\ 
             & (UTC) &   (days)      & (ks)   &  (10$^{-11}$ ergs cm$^{2}$ s$^{-1}$) & & \\ 
              \hline
      3002 & 2024-07-15T22:26:09  & 15.3 & 24.0 & 4.63$^{+0.04}_{-0.04}$  & 1.79 $\pm$ 0.01 & 822/866  \\
      3004 & 2024-07-19T00:06:09  & 18.4 & 23.7 & 4.66$^{+0.04}_{-0.03}$ & 1.72 $\pm$ 0.01 & 860/925 \\ 
      3006 & 2024-07-26T19:26:09  & 26.2 & 25.2 & 4.20$^{+0.02}_{-0.03}$ & 1.58 $\pm$ 0.01 & 925/1004 \\
      3008 & 2024-07-28T13:06:09  & 28.0 & 28.2 & 4.14$^{+0.04}_{-0.03}$ & 1.51 $\pm$ 0.01 & 980/1042 \\
    \hline
    \end{tabular}
    \caption{NuSTAR Observations of the 2024 periastron passage of PSR B1259-63. The absorption column density (n$_{\rm H}$=0.7$\times$10$^{22}$ cm$^{-2}$) was taken from \citet{Chernyakova2024}. $\tau_{\rm p}$ is the periastron time (June 30$^{th}$, 2024 or MJD = 60491.592~\citep{Chernyakova2025}). All errors are at the 90\% confidence level.}
    \label{tab:1}
\end{table}

NuSTAR quality checks for each observation of the photon event list data from both FPMs (A and B), revealed two small solar flares in observations 3004 (MJD 60510.00427) and 3006 (MJD 60517.80983), which were removed despite not contributing to the background subtracted 3-79 keV source spectra. Filtering was performed on count rate spikes and intervals of high background that were more than twice the average rate over each observation, using the NuSTAR Data Analysis Software (\textit{NuSTARDAS}) v2.1.4, as part of the \textit{HEASAOFT} 6.33.2 package~\citep{Heasarc2014}. CALDB files from August 2024 were used for the analysis. Special filtering of particle activity around the South Atlantic Anomaly was not required. After the event list data were processed, they were then analyzed using \textit{nuproducts} to create the effective area (ARF) and response matrix (RMF) files for each observation. The source counts were extracted from a circular region of radius 90 arc-seconds centered at the source position. Background counts were extracted from similarly sized circular regions with a radius of 90 arc-seconds, positioned away from the circular source region on the same chip (each detector module has four chips), avoiding the gaps between the chips and their edges. The source spectrum is above the background in the 3-79 keV band, the energy band used for the NuSTAR analysis. 

\begin{figure}[ht!]
\centering
\includegraphics[width=0.5\textwidth]{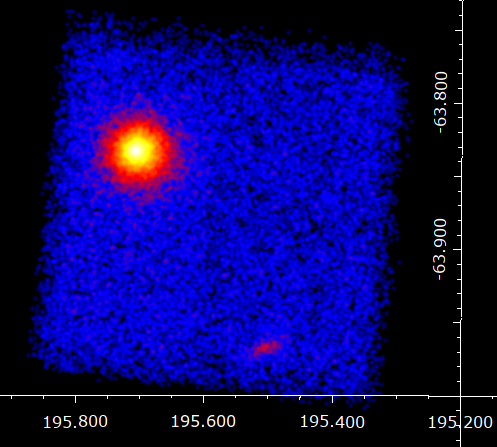}

\caption{A NuSTAR image of the region around B1259 (top left of the NuSTAR FoV) obtained during observation 3002, from FPM A. The image is smoothed with a 3 arc-second Gaussian kernel for aesthetic purposes. 2RXP J130159.6-635806 was brightest during this observation, and can be seen faintly in the bottom right of the FoV on another chip.
\label{fig:1}}
\end{figure}

Figure~\ref{fig:1} shows B1259 is clearly detected above the background in the FPM A during the first NuSTAR snapshot (3002). The variable hard X-ray source, 2RXP J130159.6-635806, appears faintly on the edge of the FoV, roughly 10 arc-minutes from B1259. This weak source was previously found to significantly contribute to measurements of the hard X-ray flux of B1259 during some of the older observations, due to previous instruments being unable to clearly resolve both sources (i.e. \cite{Chernyakova2005}). However, like the previous NuSTAR observation by \citet{Chernyakova2015}, both sources are completely resolvable. 2RXP J130159.6-635806 is much fainter during this epoch compared to the 2014 NuSTAR observations, and therefore B1259. This, and the fact that the source is on a separate NuSTAR detector chip, means counts from this source do not contaminate the source spectrum of B1259. 

\section{Results} \label{sec:res}
\subsection{NuSTAR Results}

\begin{figure}[ht!]
\centering
\includegraphics[width=0.7\textwidth]{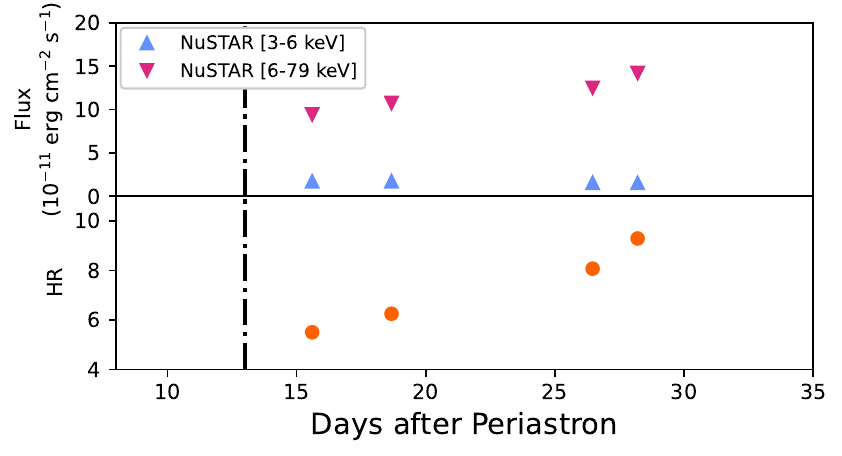}
\caption{Top: The flux for the NuSTAR observations (blue and red) plotted against days after periastron. The dashed-dotted line roughly marks the time when the pulsar crosses the stellar disk after periastron. Bottom: The Hardness Ratio (6-79~keV/3-6~keV) for each NuSTAR observation. All the data shown are for the most recent periastron passage (2024). The spectrum is observed to harden with time. The error bars on the points are smaller than the symbols.
\label{fig:2}}
\end{figure}

The NuSTAR flux for two energy bands (3-6 and 6-79~keV) is shown in Fig.~\ref{fig:2}. The data shows a decline in the 3-6 keV flux from 1.71$^{+0.02}_{-0.02}$$\times$10$^{-11}$ erg cm$^{-2}$ s$^{-1}$ around 15-18 days after periastron, to 1.53$^{+0.01}_{-0.02}$$\times$10$^{-11}$ erg cm$^{-2}$ s$^{-1}$ a week later. This is anti-correlated with the flux over the 6-79 keV band which is an order of magnitude greater and increases from 9.44$^{+0.17}_{-0.16}$$\times$10$^{-11}$ erg cm$^{-2}$ s$^{-1}$ to 14.22$^{+0.22}_{-0.26}$$\times$10$^{-11}$ erg cm$^{-2}$ s$^{-1}$ over the period of the NuSTAR observations. Taking the hardness ratio (HR), which is defined as the flux in the 6-79 keV band (hard) divided by the flux in the 3-6 keV band (soft), we see the hardening of the spectrum increasing with each subsequent NuSTAR observation, shown in the bottom panel of Fig.~\ref{fig:2}. The NuSTAR data in the 3-6 keV band are in agreement with the estimated Swift-XRT and IXPE flux~\citep{Kaaret2024}. 

\begin{figure}[ht!]
\centering
\includegraphics[width=0.7\textwidth]{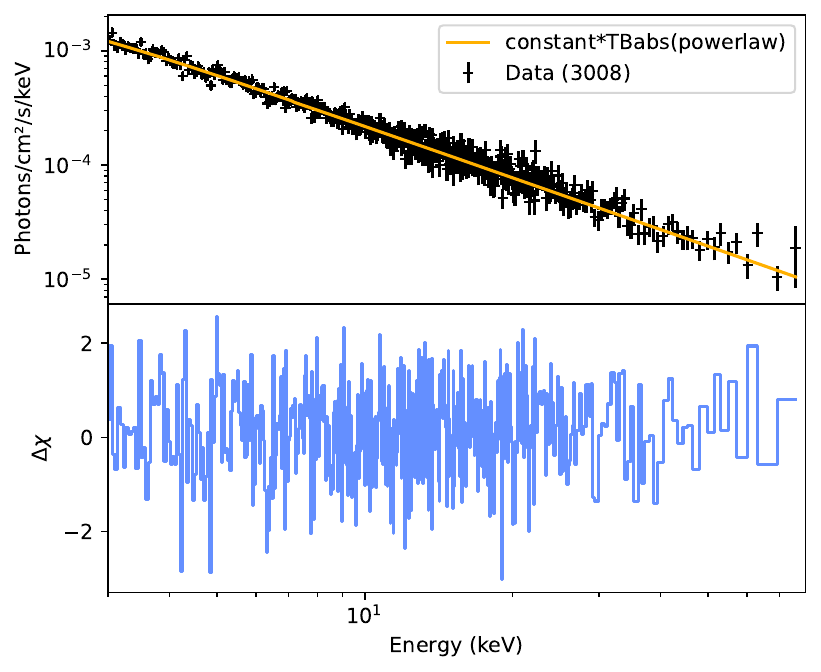}
\caption{The spectrum and model fit to the 4$^{\rm th}$, longest ($\sim$28~ks) observation. The data from this final observation was also the most energetic, with a PL model fitting the data reasonably well, with a spectral index, $\Gamma$=1.505$\pm$0.012}
\label{fig:3}
\end{figure}

We performed spectral analysis of the NuSTAR level 2 data products using Xspec~\citep{Arnaud1996}. We modeled the spectrum using an absorbed PL with the TBabs command. The absorbed column density (n$_{\rm H}$) was fixed to 7 $\times$10$^{21}$ cm$^{-2}$ as in previous studies of the same periastron passage of the source~\citep{Chernyakova2025}. These NuSTAR results are summarized in Table~\ref{tab:1} and the fitted spectrum for the fourth observation (3008) is shown in Fig.~\ref{fig:3}. The NuSTAR data is well fit by a PL model during all four observations, with a photon index that ranges from 1.79 to 1.51, in good agreement with the reported IXPE photon index of $\Gamma$ = 1.602 $\pm$ 0.013 averaged over an $\sim$800~ks observation during a similar time (July 14$^{1th}$-29$^{th}$, 2024)~\citep{Kaaret2024}. 

\subsection{IXPE and Swift-XRT Observations}
\begin{figure}[ht!]
\centering
\includegraphics[width=\textwidth]{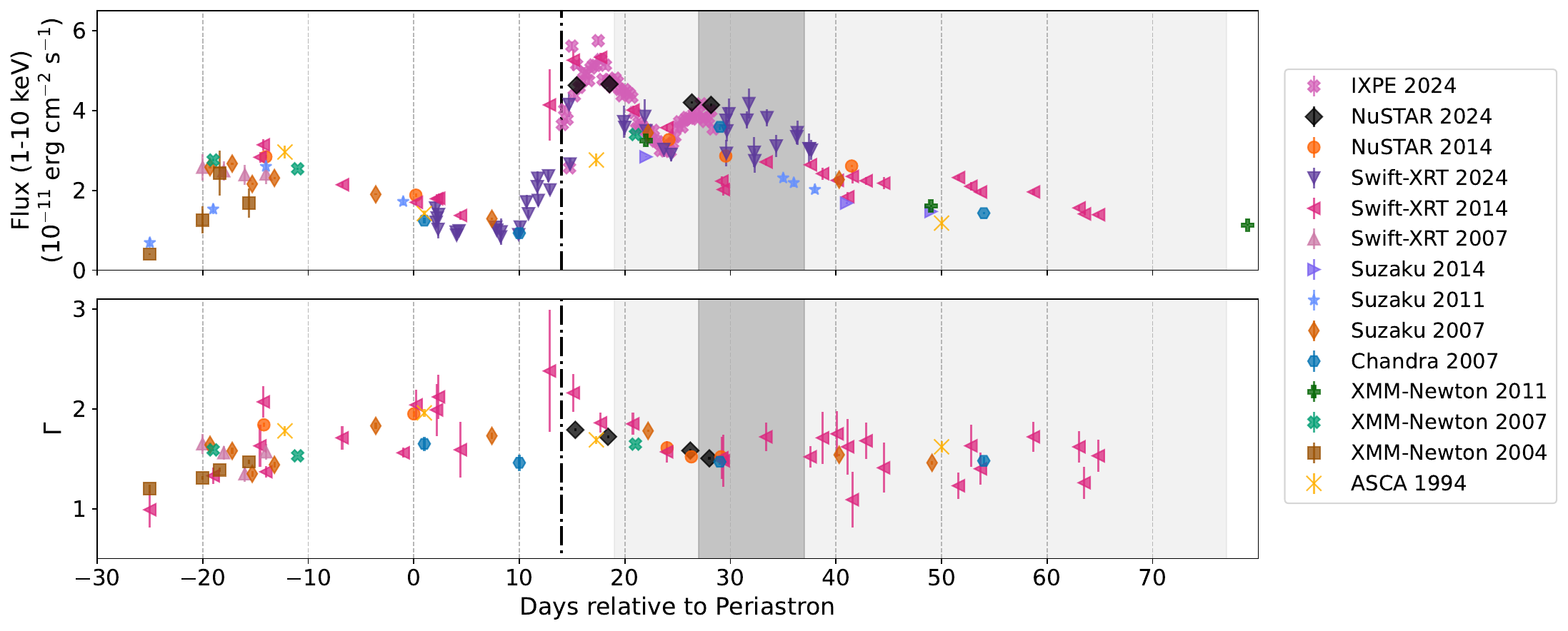}
\caption{New and historical measurements of B1259. Top: Flux over 1-10 keV. Bottom: The X-ray spectral index. All the data covers 30 days before periastron ($\tau_{p}$) and 70 days after. The dark gray block shows the first of two reported enhancements in the GeV emission during the 2024 GeV flare (light gray block, from $\tau_{p}$+19 to $\tau_{p}$+77 days as reported in \citet{Chernyakova2025}). The black dot-dashed line is the time at which the pulsar crosses the disk about 14 days after periastron~\citep{Johnston2005,Chernyakova2025}. A slow rise in the Swift-XRT flux seems to occur several days before the estimated pulsar crossing time, signaling the onset of the second bump in the X-ray flux. The flux and $\Gamma$ values are taken from \citet{Chernyakova2025,Chernyakova2014,Chernyakova2006,Uchiyama2009,Chernyakova2009,Hirayama1999,Chernyakova2015}. The most recent NuSTAR measurements (black diamond) seem in-line with previous observations around the same epoch.}
\label{fig:4}
\end{figure}

Contemporaneous IXPE observations~\citep{Kaaret2024} were taken during the NuSTAR observational campaign, starting soon after the pulsar passed through the disk, $\sim$14 days after periastron on July 14, 2024 (MJD 60505.417, a day before the first NuSTAR observation) and concluding a day after the final NuSTAR observation on July 29, 2024 (MJD 60520.163). The total source exposure collected during the entire observation by each of the three IXPE Detector Units (DUs) were 794.9, 794.5 and 794.6~ks for DU1, DU2 and DU3, respectively. More information on the IXPE results can be found in \cite{Kaaret2024}. The Swift-XRT products webpage provided by the University of Leicester was used to create the XRT~\citep{Burrows2005} data. We use 1-3.5~ks snapshot observations of ObsID 97301 (which is targeted towards B1259), with a count rate error of less than 0.1~c/s over a similar time as the NuSTAR and IXPE observations. The exposures for the NuSTAR data are roughly 25~ks each (summarized in Table~\ref{tab:1}), while the time bins for the IXPE data shown in Figs.~\ref{fig:4} and \ref{fig:5} are 19.6~ks, as in \citet{Kaaret2024}. The XRT and IXPE flux in the 1-10 keV band was estimated using WebPIMMS~\citep{PIMMS1993,Heasarc2014}, over the XRT and IXPE energy bands (0.3-10~keV and 2-8 keV, respectively), assuming an absorbed PL spectrum with $\Gamma$=1.6 and an absorption density of 7 $\times$10$^{21}$ cm$^{-2}$~\citep{Chernyakova2025,Kaaret2024}. The Swift XRT data reported here, were also presented in \citet{Kaaret2024} and \citet{Chernyakova2025}.

Figure~\ref{fig:4} shows the current observations of B1259, along with some previous observations of the source. The 2024 Swift-XRT 1-10 keV flux data in the top panel seems to show a slow rise about four days before passage commences at $\sim\tau_{p}$+14 days, seen more clearly in Fig.~\ref{fig:5}. This disk passage time is roughly the same time as previous passages~\citep{Chernyakova2025}. The 2024 Swift-XRT, IXPE and NuSTAR data in Figs.~\ref{fig:4} and~\ref{fig:5} show variation in the flux over a 3$^{rd}$ X-ray peak that occurs contemporaneously during the 2024 GeV flare (light gray box) reported in \citet{Chernyakova2025}, overlapping the first of two flux-enhanced/variable phases from $\sim\tau_{p}$+27 to 37 days (dark gray box), peaking at $\sim\tau_{p}$+30 days. The 2024 IXPE data matches the 2024 NuSTAR and Swift-XRT flux data well, resolving short-term X-ray flux variability during the 2$^{nd}$ X-ray peak, especially around 15 and 17.5 days after periastron (see Fig.~\ref{fig:5}). The 1-10 keV flux reported in Fig~\ref{fig:4} shows a noticeable jump in the flux over this energy band when compared with Swift-XRT observations of the 2021 periastron passage, which showed the absorbed flux to be lower than the 2014 and 2017 observations around 20 days after periastron as well. The flux around 40 days in the Swift-XRT data from 2024 appears to be lower than the 2021 passage, as shown in Fig.~1b of~\citet{Chernyakova2021}.

\section{Discussion} \label{sec:disc}

These 2024 NuSTAR results provide important information on the interaction region and timescales of seeding electrons at various times during the post-periastron passage epoch. The spectral index of the accelerated electrons (and therefore, the subsequent X-rays produced as a result of synchrotron emission processes), is determined by competing acceleration and energy-loss timescales~\citep{Tavani1997}. The NuSTAR observations after the 2024 periastron passage show a hardening trend over the 2$^{nd}$ bump in the 1-10~keV X-ray flux from 15 to 30 days after periastron. This can be explained by an energy-loss timescale that is longer than the acceleration timescale, which results in electrons being more efficiently accelerated to higher energies and therefore, the spectrum becoming more energetic~\citep{Tavani1997}. These energy losses after periastron are dominated by Inverse Compton (IC) emission of seed electrons from the Be star. Conversely, if the IC loss timescale becomes shorter than the acceleration timescale near periastron, then the spectrum is softer. As the pulsar moves away from the star there are fewer seed electrons, the loss timescale becomes longer, and the spectrum becomes harder.

A couple of models (i.e., coplanar and misaligned), were postulated to explore proposed mechanisms, with a shock-powered emission in the pulsar wind termination shock region~\citep{TAK94}) being the preferred model. Using this model, \citet{Tavani1997} calculated the X-ray photon index for the 2-10 keV band as a function of orbital phase (Fig. 15 of \citet{Tavani1997}). This variability with orbital phase was attributed to the inclination angle of the circumstellar disk in the binary system and its interaction with respect to the pulsar. It is the difference in this inclination angle that is responsible for the main differences between the coplanar (very small inclination angle), and misaligned (inclination angle $>$10$^{\circ}$) models. While a combination of synchrotron and IC cooling in a non-thermal, shock-powered region reproduced the observed high-energy emission from the data of previous observations of B1259 data (i.e., \citep{Bogovalov2008,Takata2012,Dubus2013,Tam2018,Chernyakova2020} and references therein), the coplanar model calculations by \citet{Tavani1997} found the predicted luminosity and spectral index to be more dominated by pure synchrotron cooling than the misaligned model, which is not in agreement with the observed data as this suggests the pulsar's magnetospheric boundaries are closer to the pulsar than the surface of the Be star. Consequently, the best model that has been used to explain the B1259 data is provided by a misaligned geometry with an inclination angle between the pulsar orbit and the Be plane constrained to 2-25$^{\circ}$~\citep{Tavani1997}. 

Fig.~\ref{fig:4} also shows the spectral index around periastron from current and historical observations. These spectral index results appear to be more complex and at odds with that predicted in \citet{Tavani1997}. A softer spectrum around periastron ($\tau_{\rm p}$-t = 0) hardens to a maximum around 15 days after periastron after which a softening trend restarts, indicating the exiting of the pulsar from the circumstellar disk. The spectral slope behavior of the first disk-crossing around 10-20 days before periastron is probably the most at odds with \citet{Tavani1997}, as it appears to get harder than what is expected at apastron. Additionally, the spectral index appears to rise/soften steadily during the first disk-crossing from $\tau_{\rm p}$-20 days to $\tau_{\rm p}$, as the flux seems to decline steadily over the same timeframe in the 1-10 keV band. This anti-correlation during the first disk-crossing could be explained by a change in the interaction cone's proximity to the pulsar resulting in the loss timescale of the electrons being longer than the acceleration timescales. However, more deeper observations during this epoch would ultimately provide higher confidence data that could provide more insightful clues of the disk environment during this first crossing, which drives the behavior observed in the data.

\begin{figure}[ht!]
\centering
\includegraphics[width=\textwidth]{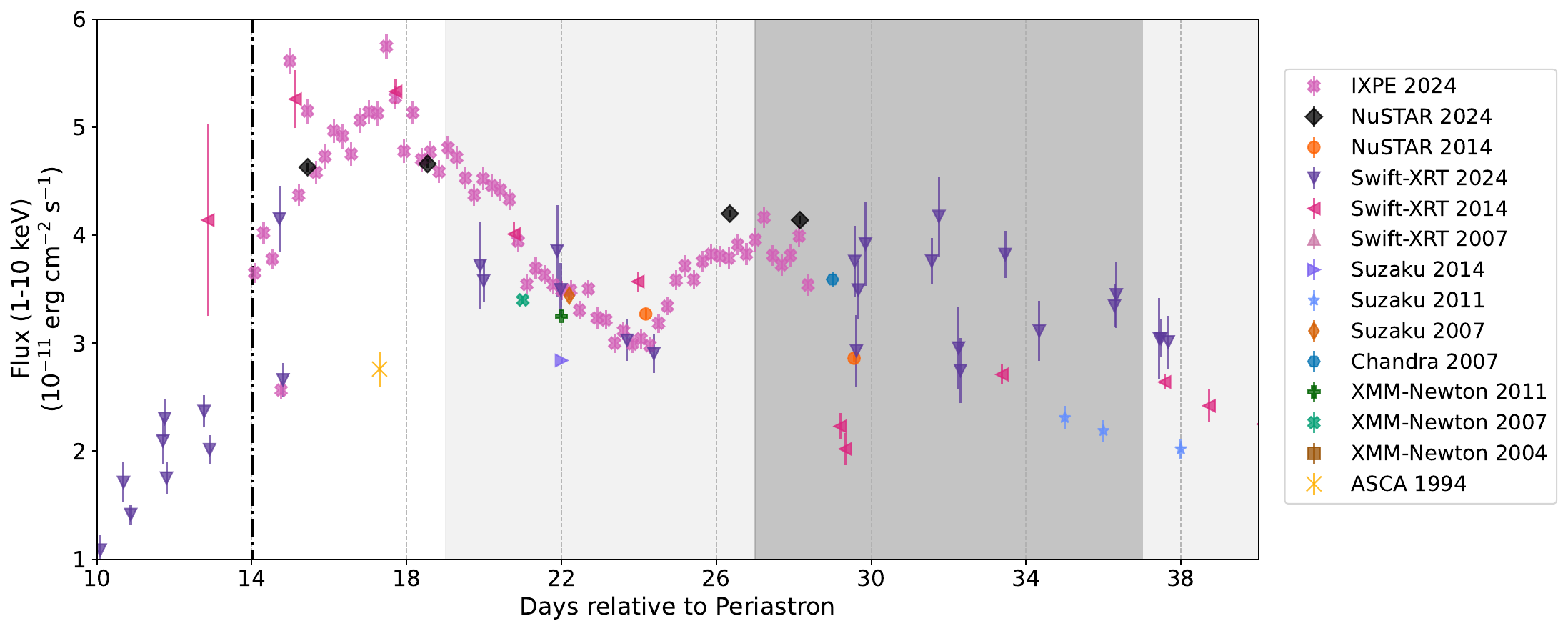}
\caption{Same as Figure 4, but focused on the X-ray flux data between 10 and 40 days after periastron ($\tau_{p}$) to show the variability in the flux during the 2$^{nd}$ and 3$^{rd}$ X-ray ``bump/peak."}
\label{fig:5}
\end{figure}

Short-term X-ray flux variability during the 2$^{nd}$ X-ray peak, is clearly resolved with the 2024 IXPE data in Fig.~\ref{fig:5}, more noticeably around 15 and 17.5 days after periastron. A significant one bin drop in the flux just after the disk crossing, followed by a couple of similarly-long flux enhancements, demonstrate the highly variable structure of the disk which has not been resolved as clearly before.

The 3$^{rd}$ X-ray flux bump is clearly seen from around 24-40 days after periastron in Fig.~\ref{fig:5}, with seemingly increased flux variability in the 2024 Swift-XRT data when compared to the other post-periastron X-ray flux bumps~\citep{Chernyakova2021,Chernyakova2025}. The variability in the X-ray flux was previously interpreted as the circumstellar disk being possibly perturbed by the pulsar, resulting in a non-uniform, clumpy disk that would move the apex of the interaction shock cone closer/further away from the pulsar surface, changing the competing timescales for energy loss and acceleration of the IC seed electrons~\citep{Chernyakova2021,Chernyakova2025}. This process may help to explain the short-scale enhancements seen over all three X-ray bump regions shown in Figs.~\ref{fig:4} and~\ref{fig:5}

The IXPE polarization data seems to possibly confirm the clumpy/perturbed disk hypothesis for the third X-ray flux maximum. The detection of polarization by IXPE from B1259~\citep{Kaaret2024} confirms a synchrotron process is responsible for the X-ray emission. Additionally, the IXPE results also provide new information on the magnetic field, suggesting the dominant magnetic field component lies perpendicular to the shock axis of symmetry. Interestingly, the polarization degree of 8.31 $\pm$ 1.45\% at a confidence level of 5.3$\sigma$~\citep{Kaaret2024}, is significantly less than the maximum possible polarization of around 70\% (assuming the spectral index of 1.6 from the IXPE observations), and greater than a random magnetic field geometry which would result in no net polarization. Therefore, the IXPE polarization degree suggests that at least $\sim$ 12\% of the magnetic field is ordered. The relatively low polarization degree (around 8\% over the entire observation), was also observed to decrease with time from around 11 $\pm$ 2\% during the beginning of the observation (2$^{nd}$ X-ray maximum) to 4$\pm$2\% during the second half of the observation (onset of 3$^{rd}$ X-ray maximum), at a confidence level of 99\%~\citep{Kaaret2024}.

In a well-ordered, highly magnetic field, we’d expect a steady enhancement of X-rays in cases where the interaction cone is consistently closer to the pulsar over longer time periods. The flux data of the 2$^{nd}$ X-ray maximum, about 17 days after the periastron time ($\tau_{p}$), could be interpreted this way, with the flaring possibly being due to local density variations in the disk and the pulsar dragging some of the disk material into it's magnetosphere during the disk passage~\citep{Chernyakova2021}. This may explain the higher polarization degree of 11\% observed by IXPE. The third X-ray flux maximum centered at $\tau_{p}$+30 days, which has more variability in the flux, may have a less-ordered magnetic field due to the disk disruption, where the interaction cone is moving more erratically toward/away from the pulsar. The polarization degree of the 2$^{nd}$ half of the IXPE observation (which includes the start of this third X-ray flux  maximum) was found to be 4 $\pm$2\%, suggesting a less-ordered magnetic field is present. Future observations of the short-scale variability during the 3$^{rd}$ bump, and of the other local X-ray flux enhancements, may provide more insight into the structure of the disk and how it changes over time. Time-resolved polarization observations around the recurring flaring times pre- and post-periastron could directly confirm whether the disk disruption alters the shock wind termination interaction region in such a way as to confidently explain the flaring nature of B1259 and the broadband emission, particularly of the the 3$^{rd}$ X-ray flux bump where it is suggested the magnetic field region becomes less ordered.

Observations from Chandra detected extended, variable emission emanating from B1259 which could be interpreted as emission from a part of the disk that had been ejected~\citep{Pavlov2015}. The partial destruction of the equatorial disk would compromise the well-organized geometry of the interacting system, including the bow-shaped contact surface of the two winds (as described in \citet{Tavani1997}). 

Multi-wavelength studies~\citep{Chernyakova2025} suggest a theoretical interpretation of the flare in which the GeV emission might possibly be due to the partial destruction of the equatorial disk of the Be star, due to the passage of the pulsar~\citep{Chernyakova2014}. While there is currently no direct evidence of this from the optical or infra-red observations, this hypothesis may motivate future searches in those data.


\begin{acknowledgments}

This work reports observations obtained with the Imaging X-ray Polarimetry Explorer (IXPE), a joint US (NASA) and Italian (ASI) mission, led by Marshall Space Flight Center (MSFC). The research uses data products provided by the IXPE Science Operations Center (MSFC), using algorithms developed by the IXPE Collaboration, and distributed by the High-Energy Astrophysics Science Archive Research Center (HEASARC). The authors acknowledge funding for this work from the IXPE and NuSTAR General Observer programs and the UK Swift Science Data Centre at the University of Leicester for the XRT data used in this work. OJR graciously acknowledges funding via grant 80NSSC24M0035 and thanks Murray Brightman for discussions of the NuSTAR data. IL was funded by the European Union ERC-2022-STG - BOOTES - 101076343. Views and opinions expressed are however those of the author(s) only and do not necessarily reflect those of the European Union or the European Research Council Executive Agency. Neither the European Union nor the granting authority can be held responsible for them. 

\end{acknowledgments}

\facilities{~NuSTAR, IXPE, Swift-XRT}

\software{\textsc{HEAsoft}~\citep{Heasarc2014}, \textsc{Xspec}~\citep{Arnaud1996}, \textsc{DS9}~\citep{SAO2000},  \textsc{AstroPy}~\citep{Astropy2013,Astropy2018,Astropy2022}}

\bibliography{NuSTAR_B1259_rev3}{}
\bibliographystyle{aasjournal}


\end{document}